\documentclass[prl,twocolumn,aps]{revtex4}

\usepackage{graphicx} % Include figure commands
\usepackage[percent]{overpic}

\newcommand{\ket}[1]{\left|#1\right\rangle}

\DeclareSymbolFont{lettersA}{U}{txmia}{m}{it}
\DeclareMathSymbol{\real}{\mathord}{lettersA}{"92}
\DeclareMathSymbol{\field}{\mathord}{lettersA}{"83}

\begin{document}

\title{Time-optimal quantum computation}

\author{Austin G. Fowler}
\affiliation{Centre for Quantum Computation and Communication
Technology, School of Physics, The University of Melbourne, Victoria
3010, Australia}

\date{\today}

\begin{abstract}
Given any quantum error correcting code permitting universal fault-tolerant quantum computation and transversal measurement of logical $X$ and $Z$, we describe how to perform time-optimal quantum computation, meaning the execution of an arbitrary Clifford circuit followed by a layer of independent $T$ gates and any necessary feedforward measurement determined corrective $S$ gates all in the time of a single physical measurement. We assume fast classical processing and classical communication, and argue the reasonableness of this assumption. This enables fault-tolerant quantum computation to be performed orders of magnitude faster than previously thought possible, with the execution time independent of the error correction strength.
\end{abstract}

\maketitle

Quantum computers promise efficient solution of many problems that are intractable on a classical computer, from factoring \cite{Shor94b}, to simulating quantum physics \cite{Lloy96}, to problems in knot theory \cite{Subr02}. A comprehensive catalogue of known quantum algorithms can be found at \cite{Jord10}. See the Supplemental Material for a basic introduction to quantum computing.

Error corrected quantum computation involves a sequence of unitary operations and measurements. All known forms of universal error corrected quantum computation use intermediate measurement results to determine future unitary operations. It is strongly believed that this is unavoidable \cite{East09,Raus12}.

Prior to this work, it was thought that error correction and the feedforward determination of future unitary operations necessarily led to substantial time overhead. Given the goal of any computing system is to solve problems as quickly as possible, this was a major shortcoming of fault-tolerant quantum computation. In this work, we show how to eliminate this time overhead given any quantum error correcting (QEC) code capable of universal fault-tolerant quantum computation and possessing the ability to fault-tolerantly measure encoded (logical) qubits in the logical $X$ or $Z$ bases through only transversal measurement of physical qubits in the appropriate basis. Almost all QEC codes of interest possess these properties, with non-Abelian codes the sole exception we are aware of \cite{Koen10,Bone12}. Even for non-Abelian codes, it is simply not known how to perform the required operations, not known for certain that the required operations cannot be performed.

\begin{figure}
\includegraphics[width=85mm]{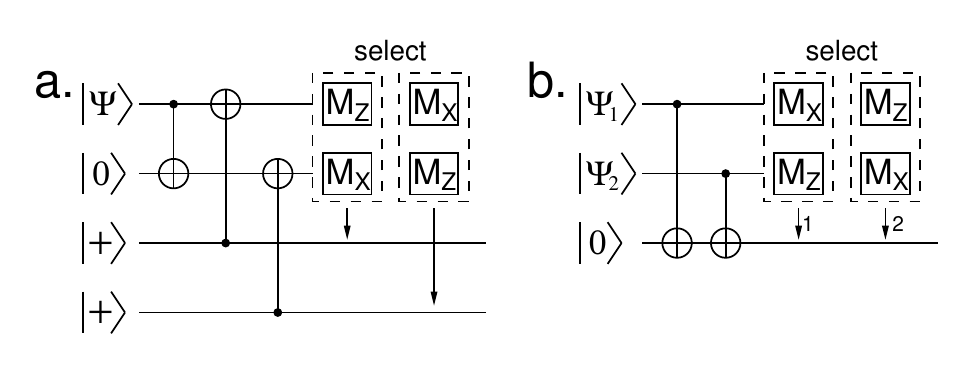}
\caption{{\bf a.}~Selective destination teleportation. If the first set of measurements is used, the bottom two qubits will be in the state $X^{M_Z}Z^{M_X}\ket{\Psi}Z^{M_X}\ket{+}$, which is a product state. If the second set of measurements is used, the output will be reversed. {\bf b.}~Selective source teleportation. If the first set of measurements is used, the bottom qubit will be in the state $X^{M_Z}Z^{M_X}\ket{\Psi_1}$. If the second set of measurements is used, $\ket{\Psi_2}$ will be teleported.}\label{teleport}
\end{figure}

Our construction is based on selective destination and selective source teleportation (Fig.~\ref{teleport}). A simple example of a circuit previously thought to require substantial time overhead is $(HT)^n$. Many processes within a quantum computer, including teleportation for data movement, introduce byproduct $X$ and $Z$ operators. Since $X$ and $T$ do not commute, with $TX=XT^\dag$, one must know after applying $T$ what byproduct operators were present to know what the result is and whether a corrective $S$ gate is required. The application of $S$, if required, cannot reasonably be delayed as $S$ does not gracefully commute through $H$ or the target qubit of $C_X$, and changes the probability distribution of an $X$ basis measurement.

\begin{figure}
\includegraphics[width=85mm]{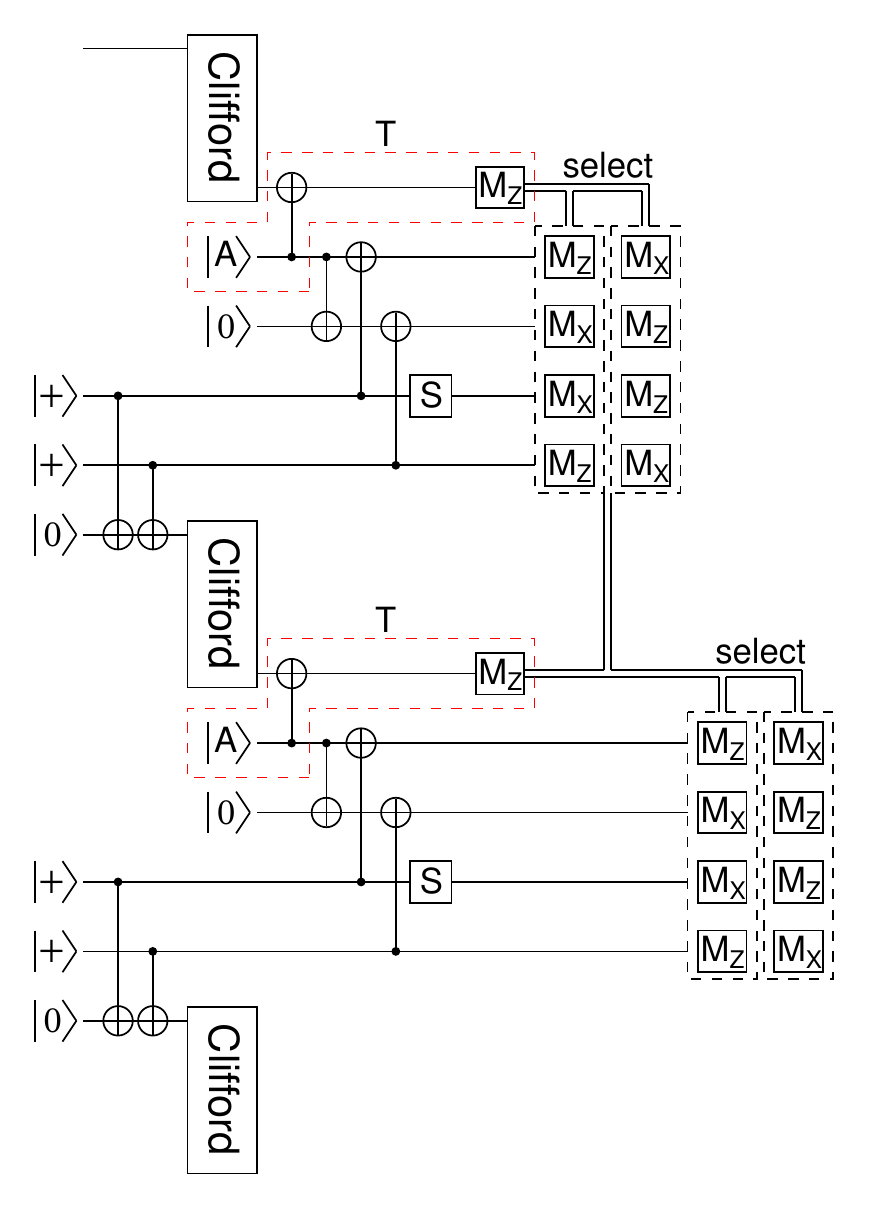}
\caption{Time-optimal quantum computation. The highlighted boxed circuit fragments labeled $T$ implement $T\ket{\Psi}$ if $M_Z=0$ and $XT^\dag\ket{\Psi}$ if $M_Z=1$. The state $\ket{A}=TH\ket{0}$ can be prepared either directly, if the underlying code supports $T$ \cite{Stea96,Alif06}, or via state distillation \cite{Brav05,Reic05}. Either way, this is an offline process that need not delay computation. In principle, all of the Clifford circuitry and all of the $T$ gates can be implemented simultaneously, however each successive decision to include or exclude a corrective $S$ gate depends on the measurements associated with the previous $S$ and $T$ gates. Double lines indicate classical signalling and processing. Assuming fast classical signalling and processing, only the physical measurement time limits the speed of quantum computation.}\label{fast}
\end{figure}

\begin{figure}
\includegraphics[width=70mm]{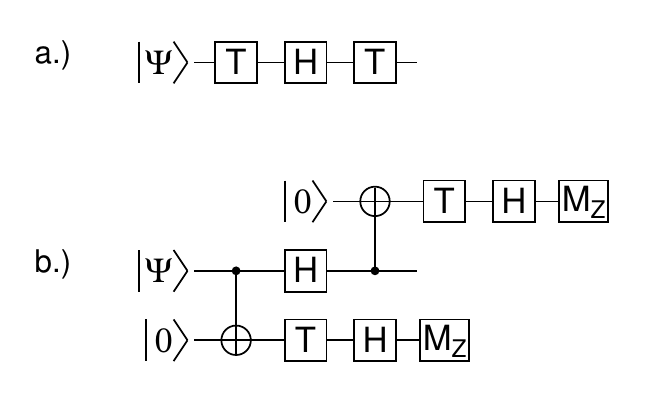}
\vspace{-5mm}
\caption{{\bf a.}~Simple quantum circuit. {\bf b.}~Same quantum circuit rewritten in an attempt to express it as a Clifford-$T$-Clifford circuit. Unfortunately, when the bottom qubit is measured, if $M_Z=1$ a byproduct $Z$ operator is introduced before the central $H$ gate. Since $HZ=XH$ and $C_X X_c=X_c X_t C_X$, where the subscripts refer to the control and target qubits, and $TX=XT^\dag$, bottom qubit $M_Z=1$ also changes the top $T$ gate into a $T^\dag$ gate, implying the two $T$ gates are not independent.}\label{dependentT}
\end{figure}

Fig.~\ref{fast} shows how to implement $(HT)^n$, including corrective $S$ gates, in a time optimal manner. Indeed, fig.~\ref{fast} shows how to implement a quantum computation composed of $n$ Clifford circuit and independent $T$ gate layers in asymptotic time $nt_M$, where $t_M$ is the physical measurement time. An example of $T$ gates that are not independent is shown in fig.~\ref{dependentT}. Care must be taken to ensure measurements do not introduce byproduct operators that modify the action of other $T$ gates in a layer.

To illustrate the utility of fig.~\ref{fast}, we shall analyze a surface code computation with and without time-optimal techniques. The surface code \cite{Brav98,Denn02,Raus07,Raus07d,Fowl12f} suppresses errors very well given only a 2-D square lattice of qubits with nearest neighbor interactions. Threshold error rates of approximately 1\% have been found in simulations \cite{Wang11,Fowl11b}, with exponential suppression of error with increasing lattice size provided all quantum gates have error rates below the threshold. Furthermore, there are strong theoretical reasons to believe the surface code is the lowest overhead code that will ever exist for a 2-D nearest neighbor architecture \cite{Fowl12h}.

Surface code quantum computation is achieved by switching off regions of error detection in a careful manner \cite{Fowl12f}. The details of how and why are not important for the purposes of this paper. A space-time region in which error detection has been turned off is called a defect. A defect that permits chains of $Z$ ($X$) errors to end undetectably on its surface is called a primal (dual) defect. The minimum number of errors required to encircle any defect or topologically nontrivially connect defects is called the distance of that surface code. The distance $d$ measures the strength of the error correction provided errors are random and independent, since long chains of errors are then exponentially unlikely.

\begin{figure}
\begin{overpic}[width=50mm]{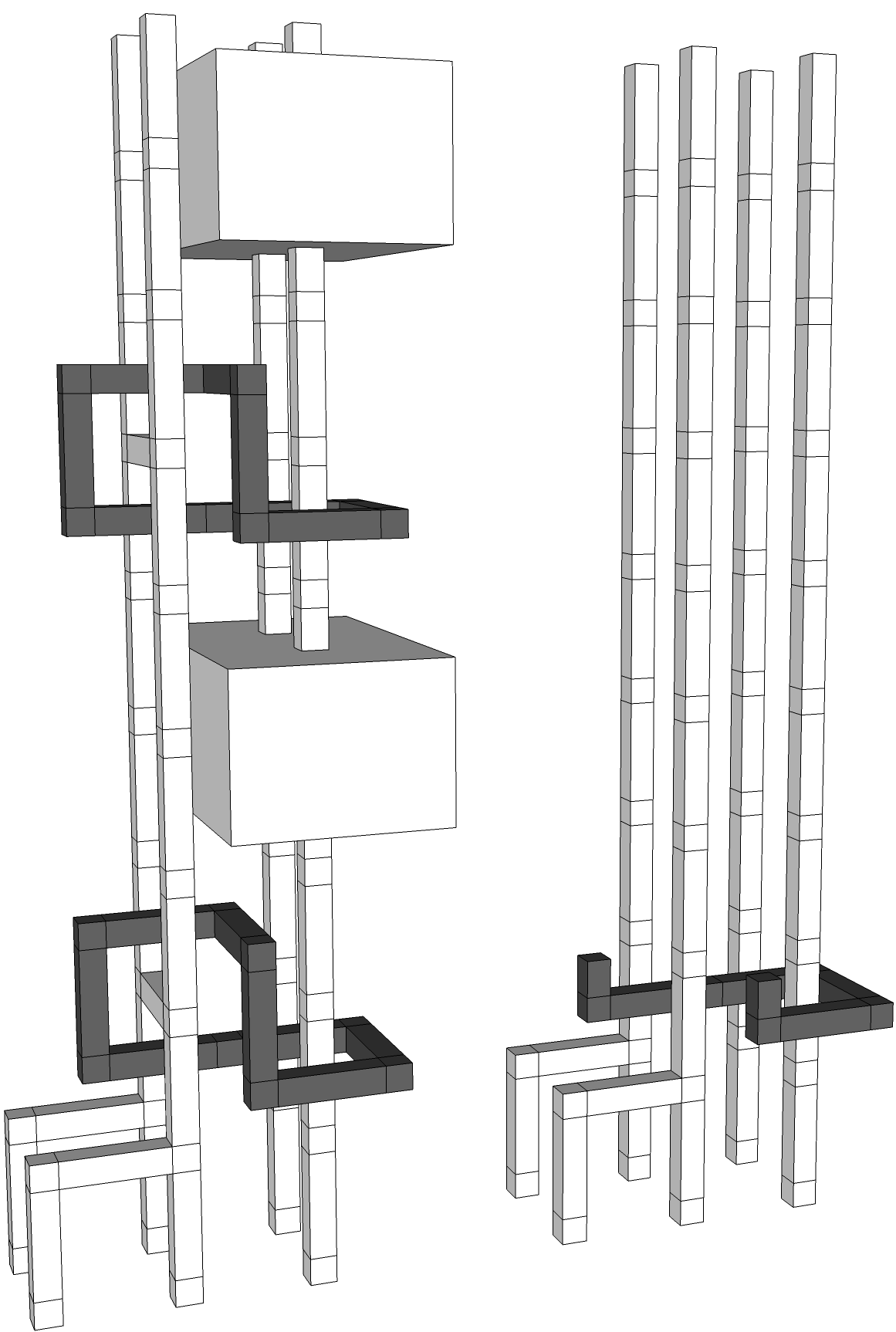}
\put(5,0){$\ket{A}$}
\put(15,1){$\ket{\psi}$}
\put(25,2){$\ket{Y}$}
\put(45,3){$\ket{A}$}
\put(53,4){$\ket{\psi}$}
\put(61,5){$\ket{Y}$}
\put(20,42){$H$}
\put(20,87){$H$}
\put(4,100){$\ket{\psi}$}
\put(14,100){$\ket{Y}$}
\put(45,98){$\ket{\psi}$}
\put(53,98){$\ket{Y}$}
\put(-2,95){(a)}
\put(36,95){(b)}
\end{overpic}
\caption{Patterns of defects corresponding to a complete surface code $T$ gate when (a) $M_Z=1$, (b) $M_Z=0$. In the latter case, we have $d$ rounds of error detection to decide whether or not to proceed with the corrective $S$ gate. Note that the incomplete section of dual defect is topologically disconnected and hence does not effect the computation.}\label{T2}
\end{figure}

\begin{figure}
\includegraphics[width=85mm]{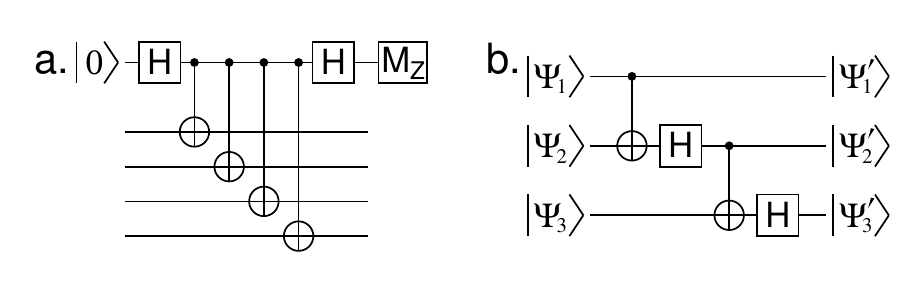}
\caption{{\bf a.}~Quantum circuit used to detect $Z$ errors in the surface code. This is the most complex and execution time limiting part of surface code error detection. {\bf b.}~A quantum circuit that at first glance require significant time to implement. Time runs left to right.}\label{Xstab}
\end{figure}

Canonical defects implementing a logical $T$ gate and its associated feedforward measurement controlled $S$ gate are shown in fig.~\ref{T2}. Given square cross-section defects of circumference $d$ and defects separated by $d$, the complete pattern has a time depth of $45d/4$, measured in rounds of error detection. A single round of error detection takes as long as the $Z$ error detection circuit of fig.~\ref{Xstab}a (the $X$ error detection circuit is shorter and all circuits are implemented in parallel). The complete $T$ gate shown in fig.~\ref{T2} would therefore take significant time to implement.

\begin{figure}
\begin{overpic}[width=45mm]{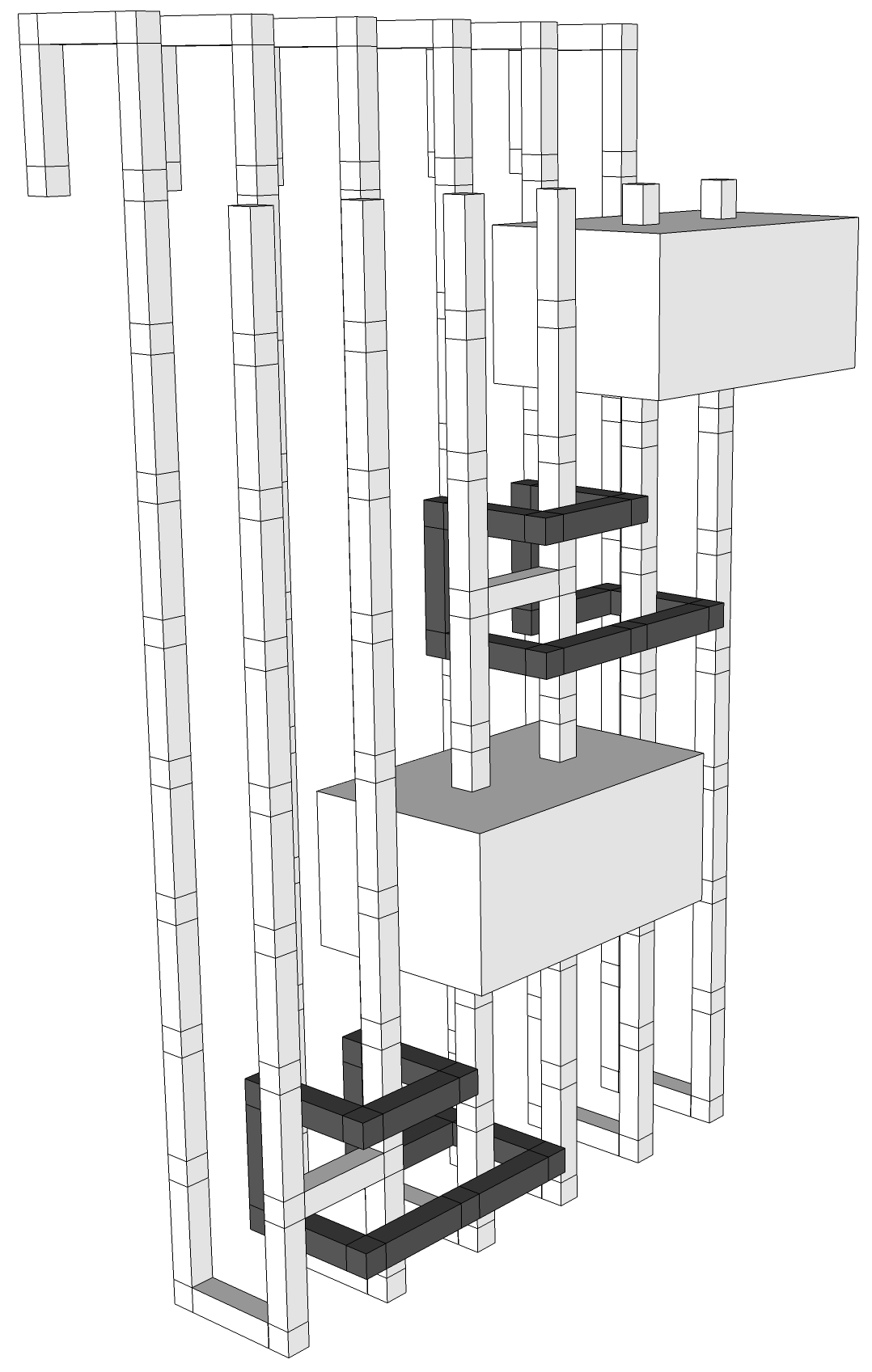}
\put(9,101){$\ket{\psi_1}$}
\put(24,101){$\ket{\psi_2}$}
\put(39,101){$\ket{\psi_3}$}
\end{overpic}
\caption{An effectively 0 time surface code implementation of the quantum circuit of fig.~\ref{Xstab}b. Time runs vertically. An arbitrary number of copies of this defect structure can be chained together without taking additional time.}\label{Hcnot_ex3d}
\end{figure}

By contrast, all other surface code gates, CNOT, $H$, $S$, initialization and measurement in the logical $X$ and $Z$ bases, do not require feedforward measurement and can therefore effectively be implemented in no time. Figs.~\ref{Xstab}b--\ref{Hcnot_ex3d} give an example of a CNOT and $H$ circuit implemented in effectively 0 time --- an arbitrary number of these circuits can be chained together sequentially and executed in constant time. Note that space-time volume is still required. This is a generic property of any QEC code executing a Clifford circuit.

Motivated by the above, it makes sense to define the $T$ depth of a quantum circuit as the minimum number of $T$ gates that need to be implemented sequentially to complete execution \cite{Amy12}. The quantum circuit $THT$ has proven minimum $T$ depth 2 irrespective of how many ancilla qubits are made available and how many additional Clifford group gates are used \cite{Seli12}. The $T$ depth, and the amount of time required to implement a $T$ gate, give the minimum execution time of the quantum circuit. Finding minimum $T$ depth versions of quantum circuits is an important and active area of research \cite{Boch12,Kliu12}. As discussed earlier, care must be taken to ensure that $T$ gates assumed to fit in a single layer are indeed independent.

\begin{figure}
\includegraphics[width=60mm]{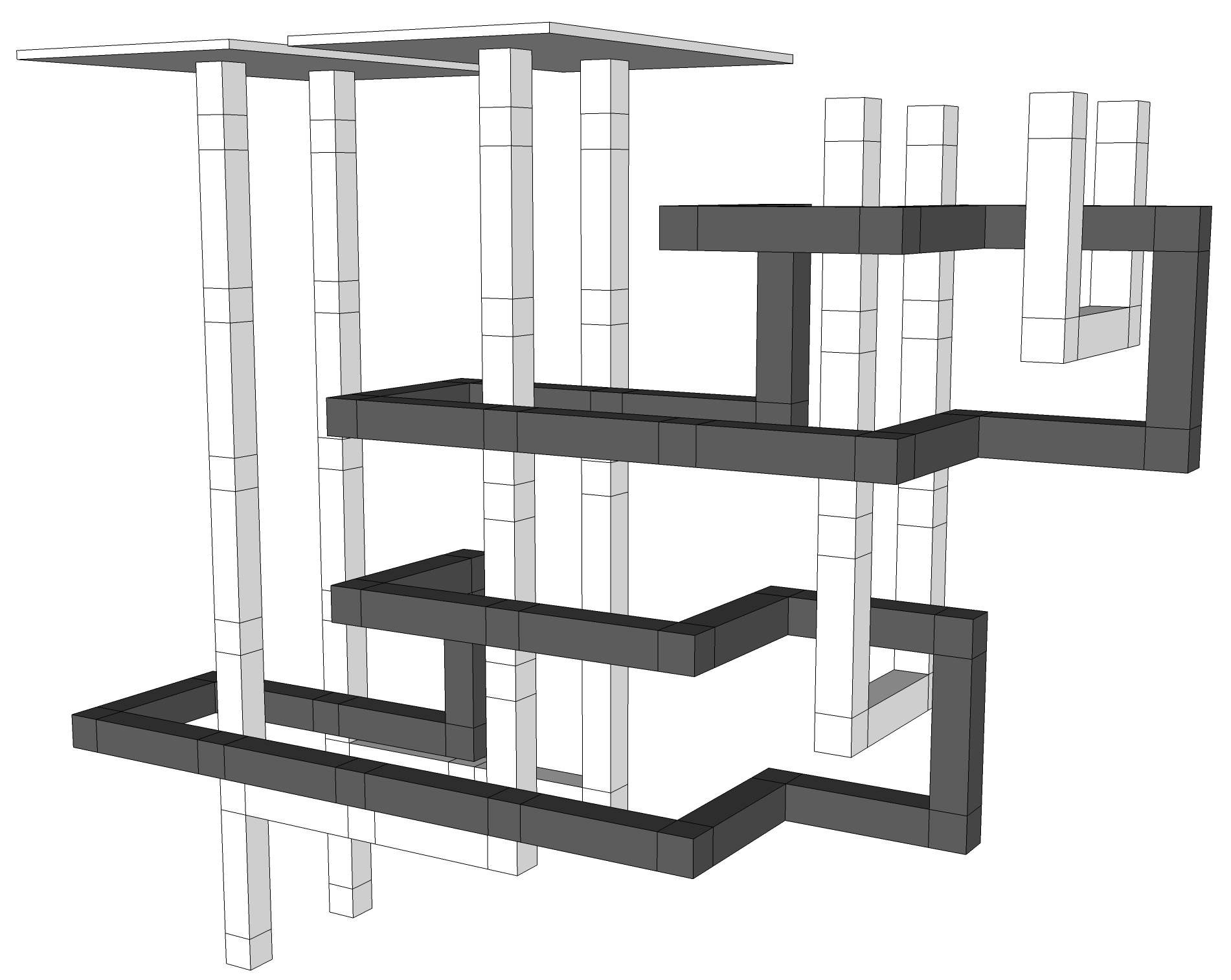}
\caption{Defect pattern implementing the selective destination teleportation of fig.~\ref{teleport}a. The underside of the caps correspond to transversal physical $X$ or $Z$ basis measurements.}\label{teleport3d}
\end{figure}

To achieve time-optimal computation, we need transversal $X$ and $Z$ basis logical measurement. Given two primal defects that collectively represent a single logical qubit, we can achieve this by measuring a region of physical qubits around each defect in either the $X$ or $Z$ basis \cite{Fowl08}. We shall represent such an operation by a flat cap on the defects. Error correction involves minimum weight perfect matching \cite{Edmo65a,Edmo65b,Fowl11b,Fowl12c}, which is a local procedure provided the error rate is below threshold. This means that by using a sufficiently large cap of characteristic dimension $d$, we can make it exponentially unlikely that measurement data outside the underside of the flat cap will be required to reliably correct the physical measurement results local to the defects that we are interested in. Fig.~\ref{teleport3d} shows a surface code implementation of selective destination teleportation. A more complex structure could clearly implement fig.~\ref{fast}.

In the surface code, overhead is currently dominated by the preparation of $\ket{A}$ states for use in $T$ gates, although this is gradually changing as new techniques emerge \cite{Brav12b,Jone12,Fowl13a}. The volume of the structure in fig.~\ref{teleport3d} is $\sim$80 (count the number of small cubes in a minimum size cuboid volume containing the structure). A slightly smaller structure is required for selective source teleportation. The volume of these structures could be substantially reduced using bridge compression techniques \cite{Fowl12h}, with a total volume of 100 conservative after compression. The volume of preparing a single $\ket{A}$ state is either approximately 200 or 600 depending on the fidelity required \cite{Fowl12h}, and the volume of the $T$ gated itself and $S$ gate correction approximately 50, so achieving time-optimal quantum computation requires a worst-case 40\% space-time volume increase. For larger problems using the larger volume state distillation structure, this additional overhead currently sits at approximately 15\%.

In the work of \cite{Fowl12f}, we estimated that $d=34$ was required to factor a 2000 bit integer using a particular superconducting architecture with measurement time 100 ns and total error detection cycle time 200 ns. Using this hardware, the canonical implementation of $T$ in fig.~\ref{T2} would take 76.5 $\mu$s versus just 100 ns (assuming fast classical processing and signalling) using the techniques summarized in fig.~\ref{fast}. This is an improvement of nearly 3 orders of magnitude. Concatenated codes in 2-D, with their extremely large time overheads for modest concatenation, would exhibit an even greater improvement.

In summary, we have described a technique enabling almost all known fault-tolerant universal quantum computation enabling QEC codes, with the sole known exception of non-Abelian topological codes, to achieve time-optimal quantum computation with modest additional quantum circuitry. Algorithm execution can asymptotically be performed in a time equal to the number of layers of independent $T$ gates times the physical measurement time. For many QEC codes, the qubit overhead is largely determined by the total number of $T$ gates. This motivates the search for low $T$ depth and count quantum circuit implementations of algorithms. Our result is fundamentally dependent on the assumption of fast classical processing and signalling, which is reasonable given the required classical processing for concatenated codes typically trivial, and that of the surface code can be performed in $O(1)$ time using a uniform array of processors on average communicating only with their nearest neighbors \cite{Fowl12c}. Nevertheless, our desire to assume that this classical processing occurs in a time negligible compared to the quantum measurement time motivates further study of the classical processing, particularly that required for block codes, including implementation in hardware. Long-term, implementation in high-speed, low power dissipation, single flux quantum superconducting technology is an attractive option \cite{Mukh11,Herr11}.

This research was conducted by the Australian Research Council Centre of Excellence for Quantum Computation and Communication Technology (project number CE110001027), with support from the US National Security Agency and the US Army Research Office under contract number W911NF-08-1-0527. Supported by the Intelligence Advanced Research Projects Activity (IARPA) via Department of Interior National Business Center contract number D11PC20166.  The U.S. Government is authorized to reproduce and distribute reprints for Governmental purposes notwithstanding any copyright annotation thereon.  Disclaimer: The views and conclusions contained herein are those of the authors and should not be interpreted as necessarily representing the official policies or endorsements, either expressed or implied, of IARPA, DoI/NBC, or the U.S. Government.

We acknowledge helpful discussions with Sergey Bravyi, Peter Selinger, Raphael Silva, and Ernesto Galvão.

\bibliography{../References}

\appendix

\section{Supplemental Material: Quantum Information}
\label{qinf}

Quantum computers manipulate quantum systems with two relatively stable quantum states that are denoted $\ket{0}$ and $\ket{1}$. These quantum systems are called qubits. Unlike classical bits, which can be either 0 or 1, qubits can be placed in arbitrary superpositions $\ket{\Psi}=\alpha\ket{0}+\beta\ket{1}$, where $\alpha,\beta\in\field$ and $|\alpha|^2 + |\beta|^2=1$. The quantities $|\alpha|^2$ and $|\beta|^2$ represent the probabilities that the qubit, if measured, will be observed to be $\ket{0}$ or $\ket{1}$, respectively.

A number of unitary matrices have specific names and symbols in quantum information. The simplest of these are the Pauli matrices, which are called the quantum gates $X$, $Y$ and $Z$. Given the definitions
\begin{equation}
\ket{0} = \left( \begin{array}{c} 1 \\ 0 \end{array} \right), \hspace{10mm}
\ket{1} = \left( \begin{array}{c} 0 \\ 1 \end{array} \right),
\end{equation}
we have
\begin{eqnarray}
\ket{\Psi} & = & \left( \begin{array}{c} \alpha \\ \beta \end{array} \right)
\end{eqnarray}
and
\begin{eqnarray}
X & = & \left( \begin{array}{cc} 0 & 1 \\ 1 & 0 \end{array} \right), \\
Y & = & \left( \begin{array}{cc} 0 & -i \\ i & 0 \end{array} \right), \\
Z & = & \left( \begin{array}{cc} 1 & 0 \\ 0 & -1 \end{array} \right).
\end{eqnarray}

The Hadamard gate $H$ is defined to be
\begin{eqnarray}
H & = & \frac{1}{\sqrt{2}}\left( \begin{array}{cc} 1 & 1 \\ 1 & -1 \end{array} \right).
\end{eqnarray}
The phase gate $S$ and so-called $\pi/8$ gate $T$ are defined to be
\begin{eqnarray}
S & = & \left( \begin{array}{cc} 1 & 0 \\ 0 & i \end{array} \right), \\
T & = & \left( \begin{array}{cc} 1 & 0 \\ 0 & i \end{array} \right).
\end{eqnarray}

Given two qubits in states $\ket{\Psi_1}=\alpha_1\ket{0}+\beta_1\ket{1}$ and $\ket{\Psi_2}=\alpha_2\ket{0}+\beta_2\ket{1}$, the state $\ket{\Psi_1}\ket{\Psi_2}$ corresponds to the outer product
\begin{eqnarray}
\ket{\Psi_1}\otimes\ket{\Psi_2} & = & \left( \begin{array}{c} \alpha_1\alpha_2 \\ \alpha_1\beta_2 \\ \beta_1\alpha_2 \\ \beta_1\beta_2 \end{array} \right).
\end{eqnarray}
The controlled-NOT (CNOT or $C_X$) gate is a two-qubit gate
\begin{eqnarray}
C_X & = & \left( \begin{array}{cccc} 1 & 0 & 0 & 0 \\ 0 & 1 & 0 & 0 \\ 0 & 0 & 0 & 1 \\ 0 & 0 & 1 & 0 \end{array} \right).
\end{eqnarray}
CNOT flips the value of the second (target) qubit if the value of the first (control) qubit is $\ket{1}$ and does nothing otherwise. The controlled-phase (CPHASE or $C_Z$) gate is defined to be
\begin{eqnarray}
C_Z & = & \left( \begin{array}{cccc} 1 & 0 & 0 & 0 \\ 0 & 1 & 0 & 0 \\ 0 & 0 & 1 & 0 \\ 0 & 0 & 0 & -1 \end{array} \right).
\end{eqnarray}

\begin{figure}
\begin{center}
\resizebox{55mm}{!}{\includegraphics{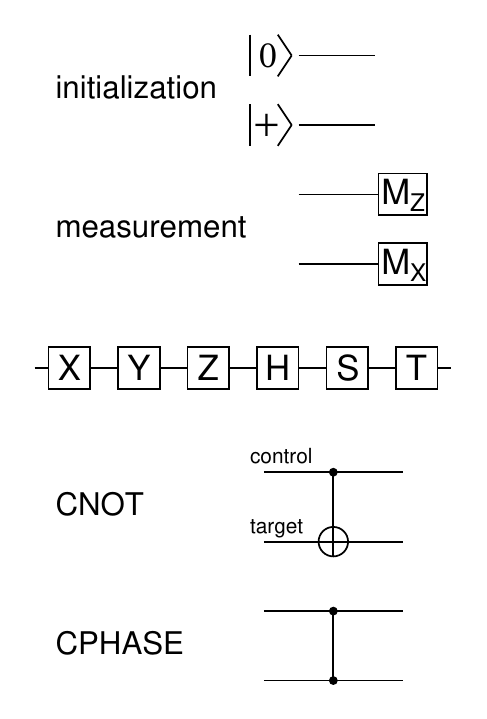}}
\end{center}
\vspace{-7mm}
\caption{Quantum circuit symbols for $\ket{0}$ and $\ket{+}$ initialization, $Z$ and $X$ basis measurement, unitary gates $X$, $Y$, $Z$, Hadamard ($H$), Phase ($S$), $\pi/8$ ($T$), CNOT ($C_X$) and CPHASE ($C_Z$). Horizontal lines represent the time evolution of a qubit. Time runs from left to right.}\label{definitions}
\end{figure}

In addition to unitary gates, the initialization gate prepares a qubit in the state $\ket{0}$. The state $H\ket{0}=\ket{+}$ is also frequently discussed as if possible to prepare directly, despite typical physical implementation as initialization to $\ket{0}$ then Hadamard. Measurement projecting an arbitrary state to $\ket{0}$ or $\ket{1}$ is called $Z$ basis measurement, and denoted by $M_Z$. Measurement in the $X$ basis is equivalent to Hadamard then $M_Z$, and is denoted by $M_X$. Quantum circuits provide a convenient notation for expressing complex sequences of quantum gates. Fig.~\ref{definitions} defines common circuit symbols.

All gates described in this section with the exception of $T$ are Clifford gates. A Clifford circuit is any quantum circuit consisting of initialization, measurement, and Clifford gates.

\end{document}